\documentclass[aps,prl, twocolumn]{revtex4}

\usepackage{graphicx}
\usepackage{amsmath,amssymb}
\usepackage{amsfonts,euscript}
\usepackage{mathrsfs}
\usepackage{subfigure}
\usepackage{boxedminipage}

\newcommand{\avg}[1]{\left\langle #1 \right\rangle}

\newcommand{\eq}[1]{Eq.~(\ref{#1})}
\newcommand{\twoEq}[2]{Eqs.~(\ref{#1}) and (\ref{#2})}

\newcommand{\fig}[1]{Fig.~\ref{#1}}

\def \be{\begin{equation}}
\def \ee{\end{equation}}
\def \bmlett{\begin{mathletters}}
\def \emlett{\end{mathletters}}

\def \FF{{\mathcal F}}
\def \HH{{\mathcal H}}
\def \LL{{\mathcal L}}
\def \NN{{\mathcal N}}

\def \Ctip {{C_\text{tip}}}
\def \Cback {{C_\text{2DEG}}}

\def \go {\gamma_0}
\def \gt  {\gamma_1}

\def \Ec {E_\text{C}}
\def \Hc {{\HH}_\text{C}}
\def \Csum {{C_\Sigma}}
\def \Cdot {{C_\Sigma}}

\def \xx {\avg{x}}
\def \uu {\avg{u}}
\def \pp {\avg{P_1}}

\def \Dsp {\Delta_{sp}}
\def \Vb {V_\text{B}}
\def \freq {\omega_0}

\begin{document}

\title{Damping of a nanomechanical oscillator
strongly coupled to a quantum dot}
\author{Steven D. Bennett}
\author{Lynda Cockins}
\author{Yoichi Miyahara}
\author{Peter Gr\"utter}
\author{Aashish A. Clerk}
\date{\today}
\affiliation{Department of Physics, McGill University, Montreal, Quebec, Canada H3A 2T8}

\begin{abstract}
We present theoretical and experimental results on
the mechanical damping of an
atomic force microscope cantilever strongly
coupled to a self-assembled InAs quantum dot.
When the cantilever
oscillation amplitude is large,
its motion
dominates the charge
dynamics of the dot which in turn leads
to nonlinear,
amplitude-dependent damping
of the cantilever.
We observe highly asymmetric lineshapes of
Coulomb blockade
peaks in the 
damping that reflect the degeneracy of 
energy levels on the dot,
in excellent agreement with our strong coupling theory.
Furthermore, we predict that
excited state spectroscopy is possible by
studying the damping versus oscillation amplitude, in analogy
to varying the amplitude of an ac gate voltage.
\end{abstract}

\maketitle

Coupling a nanomechanical object to
quantum electronics provides
a system that can be used to probe
both the mechanics and the electronics
with extreme sensitivity.
It has been predicted that the
electronics may be used to measure
the quantum nature of the mechanical object \cite{Armour02},
and the reverse---using the mechanics to measure the
quantum nature of mesoscopic
electronics---was recently
demonstrated with superconducting qubits \cite{LaHaye09}.
Electromechanical systems
that have attracted
considerable attention recently include
quantum shuttles \cite{Koenig08}, and
mechanics coupled to single 
electron transistors \cite{Knobel03,Naik06} or
tunnel junctions \cite{Flowers-Jacobs07,Poggio08}.
In most systems studied both experimentally
and theoretically, the interaction between
the electronic and mechanical components is weak.

In this paper we study
strong coupling effects, both
theoretically and experimentally, in
an electromechanical system consisting of a
quantum dot capacitively coupled to an atomic force
microscope (AFM) cantilever.
Electrons tunneling on and off the dot
effectively damp the
cantilever, and
this damping
exhibits Coulomb blockade
peaks as a function of
bias voltage similar to those well known 
in the dot conductance,
even in the limit of weak coupling
\cite{Woodside02,Zhu08,Cockins09}.
It has long been predicted that level 
degeneracy on the dot leads to 
lineshape asymmetry of
Coulomb blockade peaks in the 
conductance \cite{Beenakker91}.
Recently, we observed
corresponding temperature-dependent
peak shifts in the damping at
weak coupling \cite{Cockins09},
but the lineshape asymmetry
was far too small to be measured before now.
However, by driving the
cantilever to large oscillation 
amplitudes we enter a regime of 
strong coupling where
its motion
strongly modifies the tunneling rates
on and off the dot, and
leads to a dramatic enhancement of
the lineshape asymmetry.
This enhancement is much greater than
expected from simply extrapolating 
the weak coupling theory;
it is a non-adiabatic effect that stems
from the similarity of timescales
for dynamics of the cantilever and the dot.
Furthermore, we predict that
by measuring the damping versus 
bias voltage and oscillation amplitude,
strong coupling provides a means
to perform excited state spectroscopy 
on the dot.
Note that very different
strong coupling effects
unrelated to degeneracy were
recently reported for
a driven carbon nanotube
coupled to
an embedded dot
\cite{Steele09,Lassagne09}.

Our results show that AFM operated
at large oscillation amplitudes may
be used to study
degeneracy and level spacing,
so-called shell structure, in
quantum electronic systems.
The particular systems studied here,
self-assembled quantum dots,
are candidates for applications
such as quantum information processing,
and measuring their shell structure
has 
attracted extensive research effort
\cite{Drexler94,Miller97,Raymond04,Jung05}.
Our technique
allows the level degeneracy 
to be read off from
a single sweep of damping versus
bias voltage, and offers
the practical advantage that
non-contact AFM is a way to
address many 
dots one by one without the need for
electrical contacts.

\begin{figure}[b]
\centering
	{\includegraphics[width=0.4\textwidth]{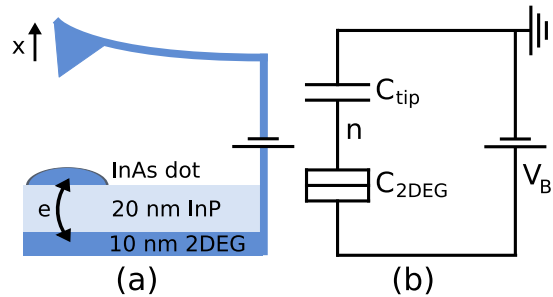}}
\caption{(a) Schematic of the setup.
Electrons tunnel on and off the dot
via a 2DEG back electrode.
(b) Equivalent circuit diagram, where
$\Ctip$ depends on $x$.}
\label{fig:system}
\end{figure}

{\it Setup}.---The mechanical oscillator is
an AFM cantilever
with resonant frequency $\freq/2\pi = 166$ kHz,
spring constant $k_0 = 48$ N/m, and
intrinsic quality factor of typically $Q_0 \sim 2 \times 10^5$.
It is driven on resonance
in self-oscillation mode at constant amplitude
and mean tip-sample gap of 19 nm \cite{Cockins09}.
The cantilever is coated with a 10~nm Ti adhesion layer
and a 20~nm Pt layer to ensure 
good electrical conductivity at low 
temperature.
All data in this paper was collected at 5K.
The sample is grown by chemical beam epitaxy, with the relevant
features being uncapped InAs dots on top of a 20~nm InP tunnel
barrier and a 10~nm InGaAs two-dimensional electron gas (2DEG)
which acts as a back electrode.
For full sample details see Ref. \cite{Cockins09}.
The bias voltage $\Vb$ is applied to the
2DEG layer, with the cantilever 
electrically grounded (see \fig{fig:system}).
The potential drop between the 2DEG and the dot
is $\alpha \Vb$, where  $\alpha = \Ctip / \Csum$ 
is extracted from the experiment and
$\Csum = \Ctip + \Cback$ is the total dot capacitance.
The dot-cantilever coupling arises through
$\Ctip$, which depends on the tip
position $x$.
Electrons tunnel between the 2DEG and the 
dot when $\Vb$ 
is sufficient to lift Coulomb blockade, while
tunneling between the
dot and cantilever is negligible due to the relatively
large distance between them.
The fluctuating charge on the dot results in both
damping and 
a resonance frequency shift $\Delta \omega$
of the cantilever; in the limit of weak coupling these
are well described by linear response \cite{ClerkBennett05}.
Here we focus on the damping, which is provided in addition
to the frequency shift by
a phase-locked loop 
frequency detector and automatic gain controller
\cite{Cockins09}.

{\it Model}.---Provided the cantilever motion is 
small compared to the
the tip-dot separation, we can assume that
$\Ctip$ depends linearly on $x$
and write the charging Hamiltonian of the dot as
\cite{Stomp05}
\begin{align}
	\Hc &= \Ec
	\left[ \left( n - \NN \right)^2 - \left(1 + \Cback/\Ctip \right) \NN^2 \right]
	\nonumber \\
	&\simeq  \HH_\text{C,0} + \Delta \HH_\text{osc} - Anx,
\label{eq:Hc}
\end{align}
where 
$n$ is the number of electrons on the dot, 
$\NN = - \Vb \Ctip/e$ is the 
dimensionless gate voltage
(or control charge),
and 
$\Ec = e^2/2\Cdot$ is the charging energy
\cite{neglectEc}.
$\HH_\text{C,0}$ is the
oscillator-independent 
part of $\Hc$,
and $ \Delta \HH_\text{osc}$,
is a constant electrostatic modification
of the oscillator potential.
Interactions between the dot and oscillator
are described by the final term
with coupling strength
$A = -\left( 2\Ec \Vb / e \right) \left( 1-\alpha \right) \partial \Ctip / \partial x$.
We stress that the strong coupling effects 
discussed here
occur despite
$\Ctip$ remaining linear in $x$;
they arise from the $x$-dependent
tunneling rates discussed below.
From \eq{eq:Hc} we see that the dot charge
exerts a force $An$ on the oscillator;
conversely, the oscillator position $x$ 
changes the energy cost of adding or removing
an electron on the dot.

We focus on the bias range where
0 or 1 extra electrons reside
on the dot, with other charge states prohibited
by Coulomb blockade;
it is simple to generalize this to the case of $n$ or
$n+1$ electrons on the dot.
The tunneling rates between the 
dot and back electrode are calculated from
Fermi's golden rule accounting for the
shell structure of the dot, i.e. the
degeneracy of single particle levels.
For a shell of degeneracy $\nu$ occupied by $n_\text{shell}$ electrons, 
there are $\eta_+ = \nu-n_\text{shell}$
ways to add an electron, and once it has been added there are
$\eta_- = n_\text{shell} +1$ ways to remove it.
The extra energy with 1 electron on the dot
(i.e. the chemical potential difference between
the dot and back electrode)
is $E(x) = 2\Ec \left( 1/2 - \NN \right) - Ax$,
which is modulated by $x$ through the 
coupling.
In the classical oscillator
limit,  
$\hbar\freq \ll k_B T$ 
\cite{QuantumCorrections,ClerkGirvin04},
this results in
$x$-dependent rates $\Gamma_+$ ($\Gamma_-$)
to add (remove) an electron,
\begin{align}
	\Gamma_+(x) &= \eta_+ \Gamma f\left[E(x)\right],
	\label{eq:rates1}
	\\
	\Gamma_-(x) &= \eta_- \Gamma \left\{1-f\left[E(x)\right] \right\},
\label{eq:rates2}
\end{align}
where
$\Gamma$ is the tunneling rate to a 
single particle state and
$f$ is the Fermi function. 
The asymmetry between adding and removing electrons
is the root of the asymmetry in Coulomb blockade peaks
\cite{Beenakker91}.

We describe the coupled system using a
master equation
for the charge on the dot combined
with a Fokker-Planck equation
for the phase space distribution of the oscillator
\cite{ArmourBlencowe04,Doiron06}.
The central objects are the
probabilities $P_0(x,u)$ and $P_1(x,u)$
to find the oscillator at position $x$ and velocity
$u$ with
0 or 1 extra electrons on the dot;
these
satisfy master equations with
$x$-dependent rates,
\begin{align}
	\partial_t P_0(x,u) &=
	\LL_0 P_0 
	+ \Gamma_-(x) P_1 - \Gamma_+(x)P_0 ,
	\label{eq:ME1} \\
	\partial_t P_1(x,u) &= 
	\LL_1 P_1
	+ \Gamma_+(x)P_0 - \Gamma_-(x) P_1,
	\label{eq:ME2}
\end{align}
where $\LL_n = 
\freq^2 \left( x - x_n - \FF/k_0 \right) \partial_u  -u\partial_x + \go \partial_u u$
describes the evolution of a driven, 
damped harmonic oscillator and
$x_n = An / k_0$ is the equilibrium position with $n$
electrons on the dot.
The damping coefficient $\gamma_0$ is 
intrinsic to the oscillator without coupling to the dot, and
$\FF$ is the external driving force.

While it is straightforward
to simulate the master
equations directly, 
we gain further insight by
focusing on the simpler dynamics of
system averages.
Following Ref. \cite{Rodrigues07},
we make the
approximation that averages of products
may be factorized into products of 
averages.
While we lose correlations contained in
\twoEq{eq:ME1}{eq:ME2},
we find from comparison
with full simulations that  
the asymmetric 
damping lineshape is still captured.
Within this approximation we
use \twoEq{eq:ME1}{eq:ME2}
to obtain coupled equations for the average
quantities,
\begin{align}
	\partial_t \xx &= \uu ,
	\label{eq:MFx} \\
	\partial_t \uu &= \freq^2
	\left( \frac{\FF + A \pp}{k_0} -  \xx \right) - \go \uu ,
	\label{eq:MFu} \\
	\partial_t \pp &= \Gamma_+\left( \xx \right) \avg{P_0} - \Gamma_- \left(\xx\right) \pp.
	\label{eq:MFp1}
\end{align}	
We seek a solution
where the cantilever oscillates
at constant amplitude $a$, 
replicating the experiment, such that
$\avg{x(t)} = a \cos{(\freq t)}$.
Ignoring the frequency shift
due to tunneling (since $\Delta \omega \ll \freq$)
and assuming that the total damping is small
($\gamma_0 + \gamma_1 \ll \freq$, justified
self-consistently),
we find that the effective, amplitude-dependent
damping due to tunneling is given
by \cite{Rodrigues07}
\begin{equation}
	\gt   = \frac{\freq^2 A}{\pi k_0 a} 
	\int_0^{2\pi/\freq} dt \sin{(\freq t)} \avg{P_1(t)},
\label{eq:dampingMF}
\end{equation}
and obtain constant amplitude oscillations for drive
$\FF = k_0/\freq^2 \left(\gamma_0 + \gamma_1\right) \avg{u}$.
\eq{eq:dampingMF} explicitly
connects the damping to the
time-varying dot charge,
$\avg{n(t)} = \avg{P_1(t)}$,
specifically the part that 
is out of phase with the cantilever position
$\avg{x(t)}$.
Note that damping arises even in the case of weak coupling,
and is  measurable using an
oscillator of sufficiently high quality factor \cite{Cockins09},
but \eq{eq:dampingMF} remains valid at strong coupling.
It also reduces our calculation of $\gt$ to
solving
\eq{eq:MFp1} for $\avg{P_1(t)}$ numerically
(inserting $\avg{x(t)}=a\cos{(\freq t)}$), a much easier
task than directly simulating \twoEq{eq:ME1}{eq:ME2}.
\begin{figure}[tb]
\centering
	{\includegraphics[width=0.45\textwidth]{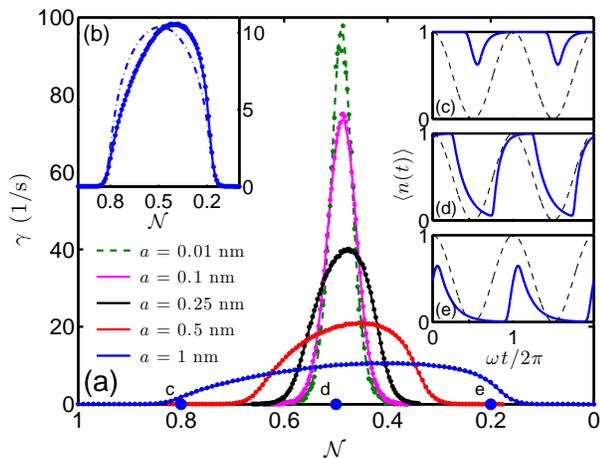}}	
\caption{({\bf a}) Calculated first Coulomb blockade damping
peak versus bias voltage
from simulation (dots) and
semi-analytic theory (solid lines) for the
oscillation amplitudes shown.
The green dashed line is the linear
response calculation.
Gate voltage is plotted in reverse for consistency with
experiment.
({\bf b}) Adiabatic approximation (dash-dotted) compared to 
semi-analytic theory (solid) and full simulation (dots)
for $a$ = 1 nm.
({\bf c-e}) Time dependence of average dot charge (solid)
for $a = 1$ nm,
at bias points marked in (a).
Cantilever position is
also shown (thin dashed) as a reference.
We took 
$2\Ec = 31$ meV, $\omega/\Gamma =1$ and
$A = 10$ meV/nm.
Other parameters are taken from the
experiment (see Setup).
}
\label{fig:simMF}
\end{figure}

Coulomb blockade 
peaks in the damping occur at charge degeneracy
points, where the dot energy is equal
with either 0 or 1 electrons and charge fluctuations
are maximal \cite{Cockins09}.
\fig{fig:simMF}(a) shows the first 
damping peak 
versus gate voltage for
several oscillation amplitudes, 
calculated both using \eq{eq:dampingMF}
and from direct simulation of \twoEq{eq:ME1}{eq:ME2}
following the approach of Ref. \cite{Doiron06}.
We assume the level structure of a 
cylindrically symmetric dot, which includes
a 2-fold degenerate $s$ shell and a
4-fold degenerate $p$ shell \cite{Kouwenhoven01}.
The simulated damping is well described 
by linear response (green dashed)
at weak coupling; note that even this peak is
very slightly asymmetric as expected.
As the oscillation amplitude is increased,
the peak becomes broadened and highly
asymmetric.
The enhanced asymmetry at strong 
coupling is completey missed in an
adiabatic approximation, where one 
assumes that the oscillator motion is 
much slower than tunneling (see \fig{fig:simMF}(b)).
This is not surprising: since $\omega \sim \Gamma$,
the cantilever can move significantly before
an electron tunnels on or off the dot.
On the other hand, the damping calculated from
our semi-analytic theory
(see \eq{eq:dampingMF})
agrees very well with the full
simulation, so we use it to understand why the 
lineshape in the damping
is so highly asymmetric at strong coupling.

{\it Asymmetric lineshape}.---The asymmetric lineshape 
of Coulomb blockade peaks 
is a result of the asymmetry 
between adding or removing
electrons to or from a degenerate
shell on the dot
(cf. \twoEq{eq:rates1}{eq:rates2}).
Consider the bias points c
and e on either side of the peak 
in \fig{fig:simMF}(a), equal distances
from its center such that the largest
amplitude oscillator
motion (broadest peak) is just large enough to swing
$\NN$ onto the charge degeneracy point.
One might guess that that a 
tunnel event near $\NN = 1/2$
is equally likely 
in both cases, but in fact 
tunneling
is twice as likely to occur
when starting from point e,
where the dot is initially empty.
This is because the rate to tunnel
onto the dot near the charge degeneracy
point is $\Gamma_+ \sim 2\Gamma f(0) = \Gamma$
(for the first peak in the 2-fold degenerate $s$ shell),
while the rate to tunnel off is only
$\Gamma_- \sim \Gamma (1-f(0)) = \Gamma/2$.
The asymmetry is apparent 
in the time dependence of $\avg{n(t)}$ 
at three bias points, shown in \fig{fig:simMF}(c-e).
Tunneling is more likely
starting from point e, and
this leads to 
increased damping.
Conversely, the situation is reversed in the second 
Coulomb blockade peak where there is only one way
to add an electron to the half full $s$ shell but two ways
to remove one once it is full.
Thus, for large amplitudes when the
asymmetry is visible, the lineshape
provides a way to read off the shell degeneracy
from a single $\Vb$ sweep: each peak is skewed
away from the center of its shell.
\begin{figure}[b]
\centering
	{\includegraphics[width=0.45\textwidth]{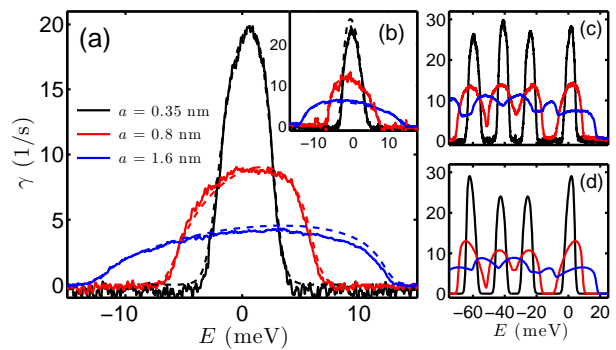}}
\caption{({\bf a}) Experiment (solid) and theory
(dashed) for the first Coulomb blockade
damping peak at three oscillation amplitudes.
We converted $e\Vb$ to $E$
using $\alpha = 0.04$ extracted at
weak coupling.
A single fit parameter value $A = 7.8$ meV/nm 
produced all three theory curves.
({\bf b}) Experiment and theory for the second peak
with $A = 9.2$ meV/nm.
The peak is skewed in the opposite direction as the first.
({\bf c}) Measured damping for
the $p$ shell; theory shown in ({\bf d}) with $A = 11$ meV/nm.
Other parameters were taken from 
experiment as described in the text.}
\label{fig:expt}
\end{figure}

While
a similar argument leads to a very slightly asymmetric lineshape
at weak coupling, the asymmetry at strong coupling
is {\it much greater} than we would expect by extrapolation.
This is because,
for sufficiently strong coupling $A$,
the change in gate voltage due to the 
oscillator motion
is greater than the thermal broadening
from the 
Fermi distribution
of electrons in the back electrode.
In other words, 
the oscillator motion dominates over temperature,
$A a \ge k_B T$.
When this is satisfied,
the harmonic distribution of the
oscillator position
$P(x)$, peaked at the turning
points of its motion, causes
the most asymmetric tunneling rates,
$\Gamma_\pm(\pm a)$ at the oscillator
extrema,
to become especially important.
The extra weighting of the most asymmetric rates
leads to the 
dramatically asymmetric lineshape at strong coupling.

Finally in our discussion of the
lineshape asymmetry, we point out
the importance of relative timescales.
For a slow oscillator, $\omega \ll \Gamma$,
the adiabatic approximation is valid since
the dot charge quickly equilibrates in response to
the slow cantilever motion.
In this case the damping is simply given by 
a weighted average of the linear
response result taken over the
oscillating gate voltage, and the lineshape asymmetry 
remains immeasurably small (see \fig{fig:simMF}(b)).
In the opposite limit,
$\omega \gg \Gamma$, the
dot charge cannot respond 
to the
rapid oscillator motion and damping is suppressed.
This can be seen from \eq{eq:dampingMF}:
for $\omega/\Gamma \gg 1$, $\avg{P_1(t)}$ 
is roughly constant over one oscillator period
and the damping becomes vanishingly small.
The case that we have focussed on
and measured
is $\omega \sim \Gamma$,
where
the interplay between the
oscillation and tunneling timescales leads to maximal
and highly asymmetric damping.

{\it Measured Damping}.---The experimentally 
measured cantilever damping 
is compared with theory in \fig{fig:expt}.
In (a) and (b) we fit the first two Coulomb blockade peaks
for three oscillation amplitudes (given in the legend)
using \eq{eq:dampingMF}.
For each charge degeneracy point, we used a single
fit parameter $A$ to fit the peak at all three
amplitudes,
obtaining the values given in the caption.
These are in good agreement with the values
obtained from our weak coupling
experiment on the same dot \cite{Cockins09}.
We took the charging energy
$2E_\text{C} = 31$ meV
and lever arm $\alpha = 0.04$
extracted at weak coupling.
In principle these parameters
may also be found by fitting the strong
coupling data directly, 
but we took advantage of our weak coupling results
as a calibration and kept them fixed.
Lastly, we point out 
that the cantilever damping is dominated by
tunneling:
here we find a peak value of
up to $\gt/\gamma_0 \sim 5$,
and at weak coupling (where the cantilever
motion is small and does not move $\NN$
off the charge degeneracy point) we 
measured as high as $\gt/\gamma_0 \sim 20$
\cite{Cockins09}.

In \fig{fig:expt}(c) and (d) we show the measured and
theoretical damping versus bias over the entire $p$ shell.
This is calculated by a straightforward extension of
our derivation of \eq{eq:dampingMF} to allow
up to four electrons to occupy the 4-fold degenerate $p$ shell.
We find good qualitative agreement even in the crude
approximation of
constant charging energy, and using single values of $A$
and $\Gamma$ over the entire shell
\cite{AandGamma}.
We took $2\Ec = 20$ meV for the $p$ shell
(estimated from the peak spacing),
and roughly aligned the peaks by adjusting
the $p$ shell level splitting
phenomenologically.
Once this was done,
a single set of parameters was used to
produce the damping spectra at all three amplitudes.
Most importantly, 
in both theory and experiment the four peaks
in the $p$-shell become five
at large amplitudes, with peaks emerging directly
{\it between} the charge degeneracy points.
This is completely consistent with our simple
theory: at large amplitudes,
the oscillator distribution $P(x)$ is peaked at its extrema and
contributions to tunneling are most important there.
Thus, for amplitudes such that 
$Aa = \Ec$, the tunneling is maximal when the
bias voltage is at the midpoint between two degeneracy points.


\begin{figure}[tb]
\centering
	{\includegraphics[width=0.45\textwidth]{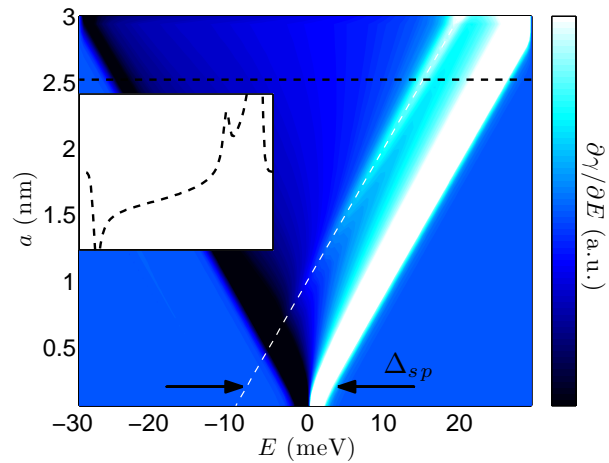}}
\caption{Differential
damping with respect to bias voltage,
plotted versus bias
(converted to $E$),
and oscillation amplitude $a$.
For sufficiently large amplitude,
a third peak appears on the line
$E = Aa - \Dsp$
(white dashed line).
Inset: cut
along black dashed line.
Parameters are the same as in \fig{fig:simMF}.}
\label{fig:levelSpec}
\end{figure}
{\it Excited state spectroscopy}.---Theoretically, 
our setup can be used to perform
level spectroscopy on the dot, for example to measure the
energy difference between the $s$ and $p$ shells, $\Dsp$.
The oscillator is directly analogous to an ac 
gate voltage on the dot \cite{Elzerman04}.
In the same way, when the oscillator motion is large enough to allow
transitions to multiple energy levels on the dot, the effective
tunneling rate increases as does the damping.
This is possible when the 
change in energy due to the oscillator is equal to the
energy spacing, or
$A a \ge \Dsp$.
At large amplitudes
we expect a jump in $\gamma$ at the bias voltage
where $E = A a - \Dsp$.
This leads to a
peak in $\partial \gamma / \partial E$
that forms a line when plotted versus
$E$ and $a$, as seen 
in \fig{fig:levelSpec}.
Measuring the slope and intercept of this line
in experiment would directly
provide $A$ and $\Dsp$.

{\it Conclusions}.---We have shown that the
oscillation amplitude is a useful 
new axis to exploit
in using mechanical measurements to
probe quantum
electronic systems.
We demonstrated our technique using 
self-assembled
quantum dots; however, its
implications extend to other quantum electronic
systems that can be placed on an insulating
surface
with a back electrode.
In particular,  it should be possible to
use AFM to study the level structure
of single molecules.
Due to the large spacing between energy levels,
these may be studied
at relatively high temperatures using
further increased oscillation amplitudes.

This work was supported 
by NSERC, FQRNT and CIFAR.

\bibliography{../../AllRefs}

\begin{thebibliography}{28}
\bibitem[{neg()}]{neglectEc}
\bibinfo{note}{We neglect the term coupling $x$ to $n^2$ since it is higher
  order in $\alpha \ll 1$.}
\bibitem[{Qua()}]{QuantumCorrections}
\bibinfo{note}{Quantum corrections to the rates occur when $\hbar\freq \sim E,
  k_B T$ (see Ref. \cite{ClerkGirvin04}), but these are negligible here.}
\bibitem[{Aan()}]{AandGamma}
\bibinfo{note}{Both $A$ and $\Gamma$ are bias dependent over the large voltage
  range of the $p$ shell.}
\end{thebibliography}
\bibliographystyle{apsrev}

\end{document}